\RequirePackage{fix-cm}
\documentclass{svjour3}                     % onecolumn (standard format)
\smartqed  % flush right qed marks, e.g. at end of proof
\usepackage{lineno,hyperref}
\usepackage{color,graphicx}
\usepackage{url}
\providecommand{\keywords}[1]{\textbf{\textit{Index terms---}} #1}
\usepackage{caption}
\usepackage{amsmath}
\usepackage{algorithm}
\usepackage{epsfig}
\usepackage{epstopdf}
\usepackage{color,graphicx}
\usepackage{lineno,hyperref}
\usepackage{url}
\usepackage{comment}
\usepackage{amsmath}
\usepackage{amssymb}
\usepackage{algpseudocode}
\usepackage{algorithmicx}
\usepackage{algorithm}
\usepackage{caption}
\usepackage{subcaption}
\algrenewcommand\textproc{}
\usepackage{epstopdf}
\usepackage{epsfig}
\usepackage{placeins}
\usepackage[T1]{fontenc}
\usepackage{txfonts}
\usepackage{array,multirow,graphicx}
 \journalname{my journal}
\newcommand{\Rmnum}[1]{\expandafter\@slowromancap\romannumeral #1@}

\begin{document}

\title{Heterogenous Networks: From small cells to 5G NR-U
}

\titlerunning{Heterogenous Networks: From small cells to 5G NR-U}        % if too long for running head

\author{Vanlin Sathya, Srikant Manas Kala, and Kalpana Naidu}


\institute{\textbf{Vanlin Sathya} \\
   Department of CSE, The University of Chicago, Illinois, USA.\\
   \email{vanlin@uchicago.edu}\\
               \textbf{Srikant Manas Kala}\\
                Mobile Computing Laboratory, Osaka University, Japan.\\
               \email{manas$\_$kala@ist.osaka-u.ac.jp}\\
               \textbf{Kalpana Naidu}\\
                Department of ECE, NIT Warangal, India\\
               \email{kalpana@nitw.ac.in }}
                %  \\

\date{Received: date / Accepted: date}

\maketitle

\begin{abstract}
{
With the exponential increase in mobile users, the mobile data demand has grown tremendously. To meet these demands, cellular operators are constantly innovating to enhance the capacity of cellular systems. Consequently, operators have been reusing the licensed spectrum ``spatially," by deploying 4G/LTE small cells (\emph{e.g.,} Femto Cells) in the past. However, despite the use of small cells, licensed spectrum will be unable to meet the consistently rising data traffic because of data-intensive applications such as augmented reality/virtual reality (AR/VR) and on-the-go high-definition video streaming. Applications such AR/VR and online gaming not only place extreme data demands on the network, but are also latency-critical. To meet the QoS guarantees, cellular operators have begun leveraging the unlicensed spectrum by coexisting with Wi-Fi in the 5 GHz band. The standardizing body 3GPP, has prescribed cellular standards for fair unlicensed coexistence with Wi-Fi, namely LTE Licensed Assisted Access (LAA), New Radio in unlicensed (NR-U), and NR in Millimeter. The rapid roll-out of LAA deployments in developed nations like the US, offers an opportunity to study and analyze the performance of unlicensed coexistence networks through real-world ground truth. Thus, this paper presents a high-level overview of past, present, and future of the research in small cell and unlicensed coexistence communication technologies.  It outlines the vision for future research work in the recently allocated unlicensed spectrum:  The 6 GHz band, where the latest Wi-Fi standard, IEEE 802.11ax, will coexist with the latest cellular technology, 5G New Radio (NR) in unlicensed.
}
\end{abstract}
%\end{frontmatter}

\section{Introduction}
Mobile networks or cellular networks have evolved rapidly over the course of the last four decades. The first generation of mobile networks (1G) were introduced towards the end of 1970s. They relied almost entirely on analog communication techniques and were primarily designed to offer voice services to the end-users. A few of the noteworthy first generation systems include the Extended Total Access Communication Systems (ETACS), the Narrowband Total Access Communication Systems (NTACS), and the Advanced Mobile Phone Service (AMPS). 1G cellular networks were followed by the second generation of mobile networks (2G) that became publicly and commercially available in the 1990s. The hallmark of 2G communication was a giant leap towards the digital mode of communication, a clear break from its analog predecessor. The switch to digital communication significantly enhanced the capabilities of the cellular systems as it offered greater network capacity, enhanced quality of voice services through intelligent speech codecs, reduced drain on the battery, and greater security. However, the landmark technological advancement was the commencement of digital data services including the Short Message Service (SMS) and Internet access over wireless cellular networks. Global System for Mobile Communications (GSM), IS-95 Code Division Multiple Access (CDMA), and IS-
136 Time Division Multiple Access (TDMA) represent the most significant 2G systems.

A few years later, General Packet Radio Service (GPRS) was introduced which employs packet switching \emph{i.e.,} data is transmitted in the form of packets. In sharp contrast, in 1G and 2G systems circuit switching was used for data transmission. A major advantage that packet switching has over circuit switching, is that the resources are reserved for an entire session and only released when the session ends (for both voice and data sessions). Packet switching became more relevant as the \emph{world wide web} or the Internet evolved and the Internet Protocol (IP) was widely adopted. These developments paved the way for an enhanced GPRS standard known as the Enhanced Data for Global Evolution (EDGE), which is an important feature of the 2.75G cellular technology stack. 

Despite, offering voice/data services and internet access of a higher quality than the earlier standards, EDGE failed to catch up with the surge in mobile data demand as the penetration of mobile phones increased and there was a tremendous increase in the number of end-users.

To meet the end-user demands, 3G cellular standards were adopted in late 1990s with the promise of higher data rates and and enhanced Quality of Service (QoS) for several data-intensive web applications \emph{viz.,} browsing, multimedia streaming, and online gaming. CDMA was one of the primary 3G technologies, W-CDMA and CDMA2000 being the most important CDMA versions.

However, the CDMA technology was not without flaws, the most important being its inherent inability to scale adequately at higher bandwidths \emph{i.e.,} at 5 MHz and beyond. Reason being, that at higher bandwidths, the data transmission rate is usually higher, leading to a shorter transmission step, which in turn exacerbates the adverse impact of multipath fading on the received signal quality. 

The demand for higher data rates continued to rise, which necessitated the research community in academia and industry to design and develop new technologies that operate well at higher bandwidths. One of the first key developments was the deployment of Orthogonal Frequency Division Multiplexing (OFDM) technology. OFDM is designed to split the data signal into multiple data streams, where each individual data-stream is allocated to a narrowband channel. Doing so ensured an increased transmission step, mitigating the adverse impact of multipath fading on the received signal. This enabled cellular providers to scale their systems for higher bandwidths by provisioning adequate number of narrowband carriers,

Multiple Input Multiple Output (MIMO) technology further enhanced the network capacity overcoming the challenges of adverse wireless environments \emph{viz.,} co-channel interference and multipath fading. In MIMO, both sender and receiver are equipped with multiple antennas, which can be used to transmit/receive data over multiple streams using special MIMO techniques. Together, MIMO and OFDM ushered in the era of high-capacity 4G cellular networks. 

\begin{table}[htb!]
\caption{Comparison of LTE and LTE-Advanced Networks~\cite{sathya2016improving,kumar2019enhancing}}
\centering	
\begin{tabular}{|p{4cm}| p{3cm}|p{3cm}|}
\hline\hline
\bfseries
\ \hspace{0.7cm}Parameter &\bfseries  LTE &\bfseries LTE-Advanced \\
\hline	
Downlink peak data rate  &300 Mbps & 1 Gbps\\
\hline
Uplink peak data rate  &75 Mbps & 500 Mbps \\
\hline
Downlink bandwidth &20 MHz & 100 MHz \\
\hline
Uplink bandwidth &20 MHz & 40 MHz\\
\hline
Bandwidth scalability & 1.4, 3, 5, 10, 15 \& 20 MHz & 20, 40, 60, 80 \& 100 MHz\\
\hline
\end{tabular}
\label{lte-a}
\end{table}

Leveraging the technological foundations of the GSM technologies, standards and specifications for 3G mobile systems were designed by the Third Generation Partnership Project (3GPP), structured as 'releases'. The vision and purpose of 3GPP is to continually develop and maintain cellular technologies and their related standards such as GSM and GPRS/EDGE, UMTS and HSPA, LTE and LTE-A. 4G or Long Term Evolution (LTE) standards were introduced by 3GPP in Release 8, with numerous technological advantages that offer enhanced network coverage and throughput, enhanced Quality of Experience (QoE), and seamless backward compatibility with legacy cellular technologies such as 2G and 3G. Further, LTE-Advanced (LTE-A) meets the specifications of LTE prescribed by International Mobile Telecommunications-Advanced (IMT-Advanced) and represents the improvements made in LTE on and after Release 10. A comparison of LTE and LTE-A networks is presented in Table~\ref{lte-a} and a summary of the technological advancements in LTE-A is presented below:
\begin{enumerate}
 \item \textbf{Carrier Aggregation:} Introduced in Release~10, the technique seeks to aggregate multiple carriers to enhance system bandwidth~\cite{alkhansa2014lte}. While, the maximum network bandwidth is 100~MHz, carrier aggregation makes it possible to achieve peak data rates of upto 1~Gbps in downlink and upto 500~Mbps in uplink. Carrier aggregation also offers several other benefits such as carrier load balancing, interference management and mitigation, QoS differentiation, and optimal heterogeneous network deployment~\cite{sathya2019auto,adam2019detection,sathya2020machine,garg2019sla,kala2019odin,sathya2020wi}. 
 Three carrier aggregation scenarios are possible depending upon the bandwidth available. They are: (a) contiguous aggregation in a single radio band, (b) non-contiguous aggregation in a single radio band,  and (c) non-contiguous aggregation in multiple radio bands. 
 
 \item \textbf{Co-ordinated Multi-point Transmission and Reception (COMP):} Proposed in Release~11, COMP makes it possible for users in the overlapping zone of multiple BSs~\cite{sathya2015femto} to be jointly scheduled. In COMP, multiple neighbouring BSs allocate the same radio resources to the UE leading to an enhanced throughput and improved QoS for the end-users at the edge of the cell. Further, dynamic coordination of transmission and reception over several BSs within the network is also made possible by COMP. Both, homogeneous networks comprising of BSs of identical types and power class, and heterogeneous networks comprising of BSs of different type and/or power classes, are supported by COMP.
 \item \textbf{Device to Device (D2D) Communication:} Generally, when two UEs communicate in the legacy cellular networks, the base station is in charge of the data plane and the control plane. ~\cite{ramamurthy2016improving,kumar2017enhancing,akilesh2016novel,ramamurthy2019dynamic}. However, in D2D communication, the communicating end-user devices data exchange directly without involving the BSs, while the control-plane signalling is overseen br the BSs. Thus, the BSs regulate the use of shared system resources to ensure optimal performance in both, regular and D2D sessions~\cite{kala2018exploring,kala2019designing}. D2D is standardized for proximity-centric communication by 3GPP in Release~12 and a typical D2D communication scenario is presented in Fig.~\ref{tradd2d}. The major challenges in D2D communication include, but are not limited to, interference mitigation and management, optimal resource allocation, session management, mobility and multi-hop network management, power control, communication security, location estimation, \emph{etc}.
 \begin{figure}
 \begin{center}
   \includegraphics[width=8 cm]{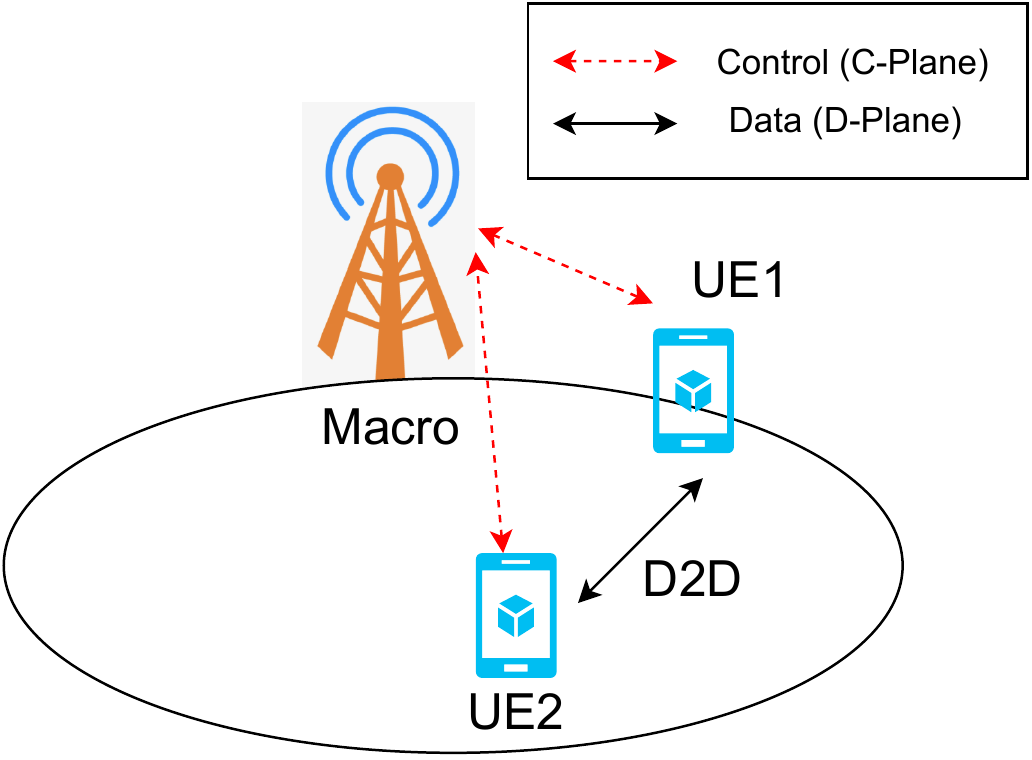}
			\caption{Typical example of D2D communication}
			\label{tradd2d}
			\end{center}
			\end{figure}
			
  \item \textbf{Dual Connectivity:}  The concept of dual connectivity was introduced by 3 GPP in release 12~\cite{agiwal2021survey,polese2017improved}, which allowed a UE to connect to both, a Macro evolved NodeB (eNB) and a small cell eNB. Further, both data and control plane communication is facilitated through both Macro and small cell eNBs. This is particularly useful for UEs located in the cell edge or in the overlapping coverage region of both type of eNBs, as it significantly enhances the throughput.
\item \textbf{LTE-WiFi Interworking ~\cite{A1,BI,A4,A3,8,PL}:} This feature was introduced by 3GPP in  Release~9. Wi-Fi, operates in the unlicensed spectrum and offers the cellular operators and opportunity to enhance network capacity by offloading mobile data from cellular networks onto Wi-Fi. ~\cite{baswade2016unlicensed,iqbal2017impact,mehrnoush2018analytical,sathya2018association,sathya2018analysis,sathya2018energy}. This is particularly true for indoor environments where cellular penetration maybe low and mobility is not a primary concern~\cite{comsnets}. Interworking of Wi-Fi and LTE ensures maximal spectral utilization, thereby improving QoS of data flows and in turn, overall network capacity. Release~13 further standardized LTE-WiFi aggregation (LWA)to facilitate optimal interworking of LTE-WiFi at the UE/eNB protocol stack level.

\item \textbf{Machine-to-Machine (M2M) Communication:} M2M communication is characterized by low power, low cost, resource constrained, and low bandwidth communication of a large number of Internet of Things (IoT) devices communicate with a remote server, primarily on the uplink~\cite{giluka2014class}. M2M communication differs from traditional UE-UE or H2H (Human-to-Human) communication with respect to nature, scope, and quantum of data traffic, types of communicating devices, the number of devices participating, and quite importantly, delay tolerance~\cite{info,W2,W3,E1,LWS11}. Consequently, the challenges in M2M are also quite different from the H2H communication paradigm. The primary challenges include excessive signaling overhead generated by a large number of IoT devices, providing support to devices with longer sleep cycles, ensuring low power communication, and facilitating reliable latency-critical small data uplink transmission. 
Since the traditional cellular networks were desgined with the vision and scope of facilitating H2H communication, 3GPP introduced solutions to the expected challenges in M2M communication in Release~11, where solutions such as RAN overload control (due to 
excess signaling~\cite{dama2016novel}) are proposed. Further, advanced technical solutions to the problems of power-constrained small data transmission for low-cost IoT devices have been specified in Release~12 and Release~13.
\end{enumerate}

\textit{Motivation for 5G and Beyond:} The growing penetration of high-end consumer devices (smartphones, tablets, etc.) running bandwidth-hungry applications (e.g., mobile multimedia streaming) has led to a commensurate surge in demand for mobile data (pegged to soar up to 77 exabytes by 2022). An anticipated second wave will result from the emerging Augmented/Virtual Reality (AR/VR) industry and, more broadly, the Internet-of-Things that will connect an unprecedented number of intelligent devices to next-generation (5G and beyond) mobile networks. These must, therefore, greatly expand their aggregate {\em network} capacity to meet this challenge. It is achieved by combining approaches, including multi-input, multi-output (MIMO) techniques, network densification (\emph{i.e.,} deploying small cells), and more efficient traffic management and radio resource allocation. The other alternate approach the operators are looking towards the Millimeter technology~\cite{wei2014key}, where there is more spectrum beyond 28 GHz frequency. In US, the operators like Verizon and AT\&T they deployed more small cell base stations in the downtown, stadium, open-park to boost the overall spectrum efficiency. But in real-time the challenges of NR-Milimeter in terms of blockage, multi-path, penetration, beam management, etc.

On the other side cheap, fast, and portable computing devices with ubiquitous wireless connectivity~\cite{giluka2014class,dama2016novel,dama2016feasible} can revolutionize the personal computing~\cite{kala2019designing} landscape by creating an opportunity to design an unprecedented array of new services and applications. Keeping the same philosophy in mind, we have focused mainly on maximizing the experienced data rate of an end-user by offering increased bandwidth through the coexistence of licensed wireless services (e.g., Long Term Evolution (LTE)) an unlicensed band~\cite{E1,sagari12,LWS22,LWS11}. To achieve the goal of throughput maximization~\cite{Cisco}, we focused on minimizing the interference and handling frequent handovers. It is done by optimally placing the small cells~\cite{ramamurthy2019dynamic,kala2018exploring} (\emph{i.e.,} a miniature base station, specifically designed to extend the data capacity, speed, and efficiency of a cellular network~\cite{lokhandwalaeai}) and controlling their emitting power in a dense small cell deployment scenario.
\begin{figure}
\begin{center}
\includegraphics[width=1\textwidth]{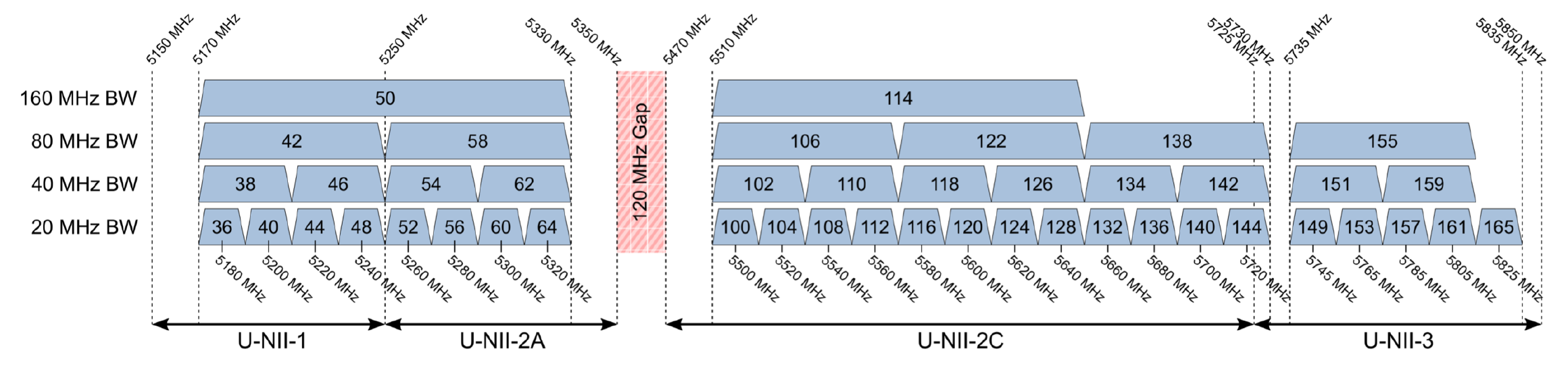}
%  \vspace{-0.5cm}
  \caption{Wi-Fi spectrum allocation in the 5 GHz Bands}
  \label{fig51}
  \end{center}
\end{figure}

{\hskip 2em} 5G LTE, which operates in the licensed band and 802.11 wireless LAN (Wi-Fi)~\cite{kala2019socio,kala2018icalm}, which operates in an unlicensed band, has some fundamental structural differences. In Wi-Fi 5 GHz, there are more spectrum available for the operators to boost the performance of the end customers as shown in Fig.~\ref{fig51}. For example, LTE control is where a base station (BS) exclusively allocates the radio resources to the users, interference due to concurrent transmission by the users handled. On the other hand, Wi-Fi~\cite{ieee} follows a distributed approach~\cite{boe,bianchi} where each user independently contests to occupy the channel, thereby concurrent transmission results in interference. The main motive of LTE/Wi-Fi coexistence in unlicensed bands for LTE users in case of very few or no Wi-Fi users. So, the research on the fair coexistence of LTE/Wi-Fi mainly focuses on the intelligent use of the unlicensed band by the LTE users to keep the Wi-Fi users unaffected so that the aim of formation of the unlicensed band remains unaltered. 

{\hskip 2em}  The standard development community has accepted two mechanisms for using the unlicensed band by LTE. These mechanisms are licensed assisted access (LAA) and LTE Unlicensed (LTE-U). LAA follows the same approach of sensing the unlicensed channel, called Listen Before Talk (LBT), as Wi-Fi. LTE-U estimates its duty cycle to access the unlicensed channel based on the various parameters such as interference, type of traffic, and load on the track.  In our recent research, we have focused on both LAA and LTE-U mechanisms. In both mechanisms, we keenly observe the various aspects of real-time, which can adversely affect the Wi-Fi users and were ignored by the existing literature. Some of the observations are:
\begin{itemize}
    \item Static channel allocation to LTE-U node in an unlicensed band
    \item Difficulty in association to the Wi-Fi access point (AP) by the Wi-Fi users due to high duty cycle (\emph{i.e.,} repeating ON and OFF intervals in the medium) in LTE-U
    \item A considerable reduction in the duration of duty cycle in LTE-U if several surrounded APs considered in its estimation.
\end{itemize}
 We used a Machine Learning approach to propose solutions based on these observations. As part of my future research plan, We want to explore the research challenges in the fair coexistence of LTE and Wi-Fi on the 6 GHz band used as an unlicensed band. Apart from that, We plan to provide Machine Learning (ML) based solutions to some of the existing problems on LTE/Wi-Fi coexistence on 5 GHz. These problems are an efficient use of high bandwidth by Wi-Fi users and optimal channel selection by both LAA BS and a Wi-Fi AP in a multi LTE operators-multi AP scenario. ML algorithms are used to closely observe the system's behavior on different conditions to make intelligent decisions.\\

The paper is organized as follows. Section 2 provides a brief overview of 4G and 5G Heterogeneous Networks. Section 3 describes the associated challenges and solutions for the past and present small cell deployments. Section 4 focuses on the future and recent NR-U small cell in 6 GHz. Finally, conclusions and future research directions are presented in Section 5.\\

\section{4G \& 5G Heterogeneous Networks (HetNets)~\cite{krishna2014dynamic,madhuri2014dynamic}:} 
\begin{figure}	
\begin{center}
			\includegraphics[width=12.5cm]{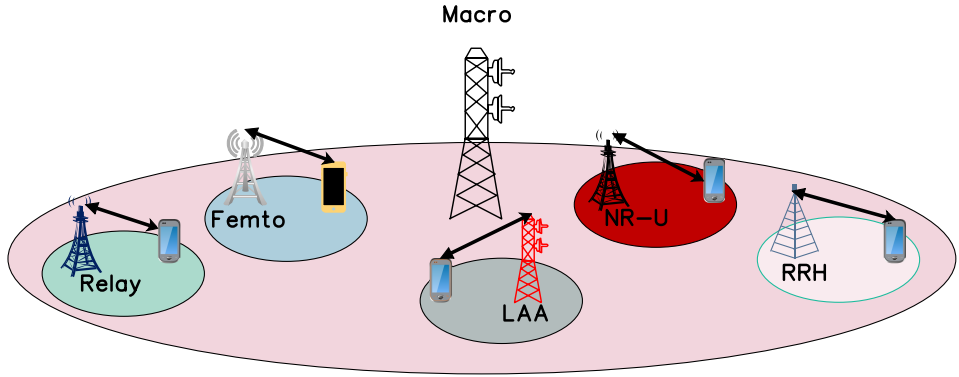}
			\caption{4G and 5G Heterogeneous Network}
			\label{hetnet}
			\end{center}
			\end{figure}

Current cellular networks comprise of end-users that generate primarily uplink traffic (\emph{e.g.,} M2M communication /IoT devices), primarily downlink   (\emph{e.g.,} streaming and web browsing), and both uplink and downlink traffic (\emph{e.g.,} gaming and social networks). Further, to offer better coverage, higher data rates, and enhanced QoS to the end-user, cellular operators have deployed low power cellular stations called small cells such as, micro cell, pico cell, Remote Radio Head (RRH), relay, and Femto cell. Together, the diversity in types and nature of mobile traffic generated by end-users and the variation in transmit powers and size of small cells, current cellular networks have lost the homogeneity of traditional networks, and are heterogeneous in nature. A typical heterogeneous network (HetNet) comprises of a Macro cell whose coverage and capacity is augmented by a variety of small cells to meet the QoS guarantees. Figure~\ref{hetnet} shows a typical example of LTE heterogeneous network. The characteristics of a variety of small cells are listed in Table~\ref{hetnettable} and a precise description of several types of small cells is presented below.
\begin{table}[htb!]
\caption{Characteristics of heterogeneous cells in 4G and 5G}
\centering	
\begin{tabular}{|p{2.5cm}| p{1.8cm}|p{2cm}|p{2.5cm}| p{1.5cm}|}
\hline\hline
\bfseries
\ \hspace{0.7cm}Technology &\bfseries  Placement &\bfseries Transmit Power  &\bfseries Backhaul Characteristic &\bfseries Number of Users\\
\hline	
Macro BS& Outdoor & 46 dBm & Dedicated wireline & 1000-2000 \\
\hline
Pico or Micro cell &Outdoor & 30 dBm & Dedicated wireline & 100-200 \\
\hline
RRH &Outdoor or Indoor &30-35 dBm & Dedicated wireline & 100-200 \\
\hline
Relay &Outdoor or Indoor & 30-35 dBm & Wireless out-of-band or in-band & 60-100 \\
\hline
Femotcell &Indoor & 20-23 dBm & Residential or enterprise broadband & 10-30 \\
\hline
LAA~\cite{WL,7} &Outdoor or Indoor & 20-23 dBm & Residential or enterprise broadband & 10-30 \\
\hline
NR-U &Outdoor or Indoor & 20-23 dBm & Residential or enterprise broadband & 10-30 \\
\hline
\end{tabular}
\label{hetnettable}
\end{table}

\begin{itemize}
\item \textbf{Pico or Micro BS:} A Pico BS transmits at 30 dBm, which is lower than the transmission power of a Macro BS. It is primarily deployed outdoors environment to offer coverage to an area of radius up to 300 m. It is connected to the Macro cell, through a dedicated $X3$ backhaul connection for efficient co-ordination as illustrated in Figure~\ref{hetnet}. 
\item \textbf{Relay:} The Relay BS serves as a repeater by boosting and re-transmitting the signal from the Macro BS, to the users connected to the Relay. Generally, cellular operators prefer to deploy Relay BSs to extend the Macro coverage and improve signal strength, especially in inaccessible terrain such as rural and hilly regions.
\item \textbf{RRH:} RRH differs from a typical BS, in that it is a radio transceiver component. It's role is limited to receiving and transmitting the In-phase Quadrature (IQ) samples. The rest of processing that usually happens at the BS is performed by a Baseband Unit (BBU) at a centralized cloud in a data center. Centralization of the core BS processing, offers the advantages of lower cost, therby lowering the CAPEX and OPEX of cellular operators.  
\item \textbf{LAA:} Unlike traditional Femtocells, LAA operates on the 5 GHz unlicensed spectrum, which does implement an LBT protocol similar to that used by Wi-Fi, with different values for parameters such as sensing threshold and transmission intervals.
\item \textbf{NR-U:} Unlike LAA cells, NR-U operates on the 6 GHz unlicensed spectrum, which implements an LBT protocol to fair share the spectrum protect the incumbent users. 
\end{itemize} 
The advantage of HetNets is as follows.
\begin{enumerate}
 \item \textit{Cell range expansion (CRE):} The possibility of increasing or decreasing the transmit power of small cell, and therefore its coverage region, depending on the traffic load significantly enhances overall network throughput~\cite{martolia2017enhancing}.
 \item \textit{Integrating Macro and small cells:} HetNets pave the way for a smooth integration of Macro and small cells that, offering dual connectivity to each device (\emph{i.e.,} a Macro and a small cell), improving the end-user QoS experience.
 \item \textit{Self organizing network (SON):} Given the large number of Macro and small cells deployed in a HetNet, to ensure seamless configuration and negligible down-times, SON capabilities are provided to every small cell within the HetNet. This not only reduces manual intervention required in installation and configuration process, but also optimizes the network performance automatically by reducing collisions at the UE during cell-selection. The SON paradigm of HetNets simplifies the OAM (operation, administration, management) aspects of complex and dense network.
\end{enumerate}

\subsection{LTE Femtocell Networks}\label{sec:LTEFemNet}
\par The existing Macro Base Stations (MBS) cannot satisfy mobile users because of most users' huge data demand and indoor locality. Reports by Cisco tell that 70\% of the traffic is generated in indoor environments such as homes, enterprise buildings, and hotspots. Hence, mobile operators must improve coverage and capacity`\cite{sathya2013dynamic} of indoor environments. But the basic problem with the existing MBS (or small outdoor cells with shorter coverage) is that they can only boost data rates of Outdoor User pieces of equipment ($OUEs$). But, they cannot do the same for Indoor User pieces of equipment ($IUEs$) because it is difficult for electromagnetic signals to penetrate through walls and floors. Owing to numerous obstacles in the communication path between MBS and $IUEs$ inside the building, radio signals attenuate faster with an increase in the distance. Thus, $IUEs$ receive low signal strength (\emph{i.e.,} Signal-to-Noise Ratio, SNR) compared to outdoor users. Hence, mobile operators must improve coverage and capacity in indoor environments.

 \par As a solution, Femtocells are being deployed by both operators and end customers. Femtocell is a low-cost, low-power consuming cellular base station that operates only in a licensed spectrum and designed for outdoor and indoor communication. The range of Femtocell is 100-150 meters for enterprise environments consuming 100 mW power. A home-based Femto (HeNB) can serve 4-5 users, whereas an office-based Femto can serve a maximum of 64 users. Each Femto requires a backhaul connection to the evolved packet core (EPC). Advantages of using Femtos are described as follows: \\ \\
\noindent \textbf{Operator Advantages :}
\begin{enumerate}

\item The operator can increase the network capacity.
\item The operator can reduce Operational expenditure (OPEX) and Capitational expenditure (CAPEX).
\item The operator can reduce Backhaul cost.
\item The operator can reduce Traffic overload on MBSs.
\end{enumerate}
\noindent \textbf{User Advantages :}
\begin{enumerate}
\item Improved Quality of Experience (QoE).
\item Improved energy efficiency/battery life.
\end{enumerate}

\subsection{Architecture of Indoor LTE Femto cells}\label{sec:LTEindoorFem}
In the LTE HetNet system's architecture, Femtos are deployed inside the building and connected to a Femto Gateway (F-GW) over the S1 interface. F-GW is mainly used to reduce the load on MME. It acts as a virtual core network to Femtos. The F-GW gets assigned with an eNB ID, and thus F-GW is considered another eNB by the MME. The X2 interface~\cite{X1,X2,X3,X4,X5} is introduced between Femtos of enterprise Femtocell networks to avoid inter-cell interference and directly route the data and signaling messages among Femtos, thereby reducing the load on LTE core network and offering better coordination among Femtos.   
\subsection{Access Modes in Femto}
Since Femtos~\cite{vanlin2013dynamic} are deployed for offering high data rates to indoor (paid) users in enterprise and residential buildings, each Femto is configured with a list of subscribers called Subscriber Group (SG) such that only the users in the SG can access the Femto. The users not belonging to this list are called Non-SG (NSG), and they may not be served by the Femto even when they are close to the Femto. Following access modes are defined for Femtos:
\begin{itemize}
\item \textbf{Open access:} The open-access mode allows all users (\emph{i.e.,} SG \& NSG) to access the Femto without any restriction.
\item \textbf{Closed access :} The fast-access mode permits only authorized users (\emph{i.e.,} SG) to access the Femto.
\item \textbf{Hybrid access:} The hybrid access~\cite{ghosh2017novel} is the combination of both open and closed access. It allows all users (\emph{i.e., SG \& NSG}) by providing preferential access for SG users over NSG users.
\end{itemize}

\subsection{LTE Licensed Assisted Access}
3GPP specifies LTE-LAA in Release 13 adopted the LBT approach for coexistence with Wi-Fi and supported only DL transmissions in the unlicensed band:  a secondary cell (sCell) aggregated with a licensed primary cell (pCell). Enhanced LAA (eLAA), as specified in Release 14, supports UL operation in the unlicensed band. However, the legacy LTE UL schedule continued to be used in eLAA, thus increasing the processing delay in scheduling grants due to LBT procedures. Hence, in April 2017, 3GPP started the "further eLAA" (FeLAA) working group (in Release 15) to improve LAA DL and UL performance through enhanced support for autonomous UL transmissions.  In the proposed FeLAA, a UL transmission ought to receive a grant from the eNB before the transmission, which solves the constraint imposed by the legacy eLAA. Most of these features have not been tested in the field before deployment. As LTE-LAA  deployments are being rolled out in major cities in the US, they offer an opportunity for real-world testing.

\subsection{5G Small Cell in Unlicensed Deployment}
The coexistence of small-cell LTE-U~\cite{A1,BI,A4,A3,8,PL,cts,blank} and Wi-Fi networks in unlicensed bands at 5 GHz is a topic of active interest~\cite{baswade2016unlicensed,iqbal2017impact}, primarily driven by industry groups affiliated with the two (cellular and Wi-Fi) segments. In contrast, there is a body of analytical work~\cite{info,W2,W3,3} exploring the coexistence of LTE-U and Wi-Fi, our focus in this project has been on real-time measurements ~\cite{xu2020understanding,narayanan2020lumos5g} and real-time deployment aspects of such coexisting networks. Coexistence is a topic that has seen little traction in the existing literature. As per the scope of this project, we actively design, analyze, and implement wireless network algorithms in simulation (using ns-3), in a real-time National Instrument (NI) unlicensed coexistence test-bed~\cite{mehrnoush2018analytical,mehrnoush2018fairness} and also conduct measurements and analysis on recently deployed LAA in the Chicago area.

In our previous work~\cite{mehrnoush2018analytical,mehrnoush2018fairness}, we investigated various aspects of coexistence between the two principle variants of LTE in the unlicensed bands, LTE-U and LAA,  and Wi-Fi~\cite{wilhelmi2019potential}. For LTE-U, we analyzed the effect of the LTE-U duty cycle~\cite{N1,E2,E3,3GPP} on the performance of the association of Wi-Fi. We demonstrated~\cite{sathya2019auto,adam2019detection,sathya2020machine,dziedzic2020machine} that using a high duty cycle adversely impacted Wi-Fi's ability to access the channel due to the Wi-Fi beacon transmission process's disruption. Therefore, we recommended using a lower duty cycle even if there was no Wi-Fi present in the channel to enable fair access to a new Wi-Fi access point that may wish to use the channel. We also developed machine-learning-based algorithms to determine the number of Wi-Fi APs on the air to enable an appropriate duty cycle setting. However, since LTE-U is not considered for wide deployment by industry, we switched our attention to evaluating LAA coexistence.

We have made tremendous progress toward understanding Wi-Fi and LAA's coexistence behavior in our previous work's unlicensed bands. Our theoretical analysis, corroborated by detailed system-level simulations using ns-3 and use of software-defined-radios, demonstrates that coexistence is improved substantially when the two systems treat each other symmetrically. That is when Wi-Fi and LAA use the same detection threshold to defer to each other. In our recent work~\cite{sathya2020measurement}, we added a new dimension by performing detailed measurements of deployed LAA networks by the three major carriers, Verizon, AT\&T, and T-Mobile, in Chicago. We conducted these measurements using off-the-shelf and custom-designed apps to extract detailed network information via APIs on Android smartphones. We believe this to be the first such exercise in academia and the measurements revealed several interesting new directions, which we will continue to research. Two such topics are: (i) Though most LAA deployments are outdoors and Wi-Fi's are indoors, the client devices that connect to these networks can be used outdoor/indoor; this results in hidden-node scenarios worse by the fact that two systems do not decode each other's signals (ii) Most academic analyses have focused on coexistence in a single 20 MHz channel, Whereas our measurements reveal that LAA usually aggregates three unlicensed channels, therefore increasing Wi-Fi's impact. These results have been presented to the industry as well and been incorporated into recommendations by Cisco.

 \begin{figure}
 \begin{center}
   \includegraphics[width=8 cm]{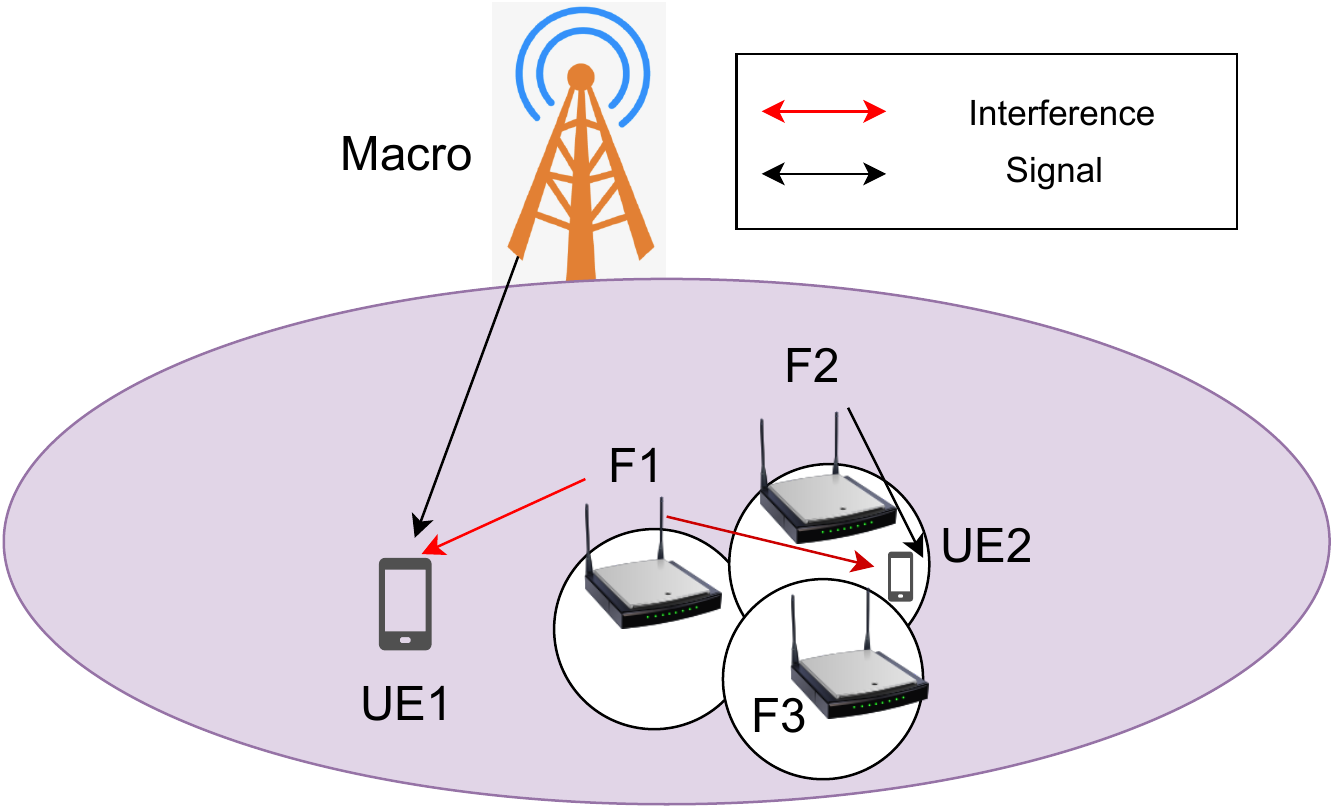}
			\caption{Co-tier and Cross-tier Interference in HetNet Small Cell Network}
			\label{cotier}
			\end{center}
			\end{figure}
			
\section{Small Cell deployments: History, Associated Challenges, and Existing Solutions}
 %include related work in here

This section discusses the various challenges and problems faced while deploying small cells in LTE HetNets. It also discuses the existing solutions to address them which primarily focus on enhancing the overall system performance. The major challenges are outlined below:\\
\begin{enumerate}
\item Placement of Small Cells in 4G LTE\\
Femto cells have been deployed at a large scale in enterprise/office settings and the main constraints that operators face are a lack of space and low power consumption ~\cite{tahalani2014optimal,sathya2015joint,tahalani2013optimal}. Due to spatial constraints, operators often resort to a sub-optimal arbitrary deployment of Femto cells. This causes coverage gaps and increases the number of Femtos required, leading to increased CAPEX and OPEX. Moreover, arbitrary placement involves human resource expenses incurred in field testing. A feasible solution to these problems is to ensure an optimal placement of Femto cells~\cite{lokhandwala2014phantom,optimal}. An optimal placement guarantees a robust SINR and high network capacity while ensuring that there are no coverage holes within the building ~\cite{tahalani2014optimal,sathya2014placement,sathya2020small}.
However, there are additional challenges that need to be addressed. Solutions proposing optimal deployment of single Femto cell are often not scalable to offer coverage to entire enterprise/office buildings. Further, proposed placement strategies often fail to consider architectural parameters of the built environment such as  wall-thickness and dynamic power transmission. Femtos may also inadvertently create coverage holes for indoor users when the operate in the closed access mode and there are non-subscriber UEs in close proximity. Consequently, some solutions have been proposed to minimize coverage gaps through efficient Femto placement and power control strategies \cite{lokhandwala2015phantom,sathya2013efficient,ramamurthy2015energy}.\\ 

In a HetNet, like most wireless networks, the dominant factor governing the system throughput is the interference. Two types of interference affect a HetNet system \emph{viz.,} interference between two or more Femtos, and interference between a Macro and a Femto. A detailed description of each is provided below:\\
\begin{enumerate}
\item \textbf{Co-tier Interference:} Reusing spectrum for maximal utilization invariably leads to  interference from the neighboring small cells. This phenomenon is called co-tier interference~\cite{sathya2016improving,sathya2016handover,sathya2020raptap}. For example, UE2 is getting served by the Femto BS (F2), but it is receiving interference from the neighboring Femto BSs (F1) as shown in Fig.~\ref{cotier}. The conventional approach to mitigating co-tier interference in a cellular network is through a mechanism known as Inter-Cell Interference Coordination (ICIC). The ICIC approach allows all BSs to communicate cooperatively using the X2 interface. While it succeeds in ensuring efficient allocation of RBs to users at the cell-edge, it leads to a significant increase in the signaling overhead.
\item \textbf{Cross-tier Interference:} This is the interference caused by transmission conflicts between a Macro BS and a small cell ~\cite{sathya2015femto,sathya2016handover,sathya2016maximizing}. For example, UE1 is getting served by the Macro BS, but it is receiving interference from small cells (\emph{i.e.,} Femtocell F1 as shown in Fig.~\ref{cotier}). The enhanced ICIC (eICIC) approach is the conventional solution to the problem of cross-tier interference~\cite{giluka2016handovers,giluka2016leveraging,sathya2020qos}. It mitigates the transmission conflicts between an MBS and a Femto BS by muting a few sub-frames (Almost Blank Sub-frame) in MBS when the Femto BS is transmitting. Doing so, alleviates the interference and enhances HetNet system capacity.\\
\end{enumerate}
The spatial reuse of the spectrum enhances efficiency and capacity of a HetNet system, but the co-tier and cross-tier interference cused by obstacles inside buildings are unintended consequences that adversely affect the signal strength and throughput of indoor devices.  In the outdoor setting, in a building under the coverage of LTE HetNet Macro BSs, signal leaks occur around its edges/corners, creating a High Interference Zone (HIZone) in its vicinity. This leads to cross-tier interference and causes similar performance bottlenecks for $OUEs$. However, the challenge of cross-tier interference for HIZone UEs ($HIZUEs$) has not been adequately addressed dynamically \emph{i.e.,} by considering the UE occupancy level within the HIZone. However, studies indicate that an active power control strategy at the Femtos is very likely to reduce the cross-tier interference in the HIZone.\\
\item Scheduling or Radio Resource Allocation \\ \\
Femtos deployed in enterprise and residential buildings guarantee high bandwidth services to a specific list of paid indoor users/subscribers called the Subscribers Group (SG)~\cite{ghosh2017novel}. Each Femto allows access to users only in the SG and the indoor users not in the list are known as Non-subscriber Group (NSG), which are served by the MBSSs, regardless to their proximity to a Femto. This subscriber-only access is known as \textit{closed access}. In sharp contrast, Femtos configured to operate in the \textit{open access} mode do not discriminate between SG and NSG users, and allow access to all indoor UEs depending on their proximity to the Femto. The pitfall is that, the open access mechanism may be unable to meet the QoS guarantees for all indoor LTE/5G users during the peak-traffic hours.  
%Hybrid access mechanism integrates the principles of both closed and open access mechanisms. It lets some NSG users connect with Hybrid Access Femtocells (HAFs) and share its radio resources and SG users. This mechanism provides a trade-off between maximizing overall HetNet capacity and maximizing the throughput of SG users. Figure~\ref{figoac} shows an LTE HetNet system comprising one Macrocell and three Femtocells: one OAF, one CAF, and one HAF. Here the OAF is serving both SG and NSG users located in its coverage area. In the CAF case, all of the NSG users located in the CAF's coverage area are forced to connect with the MBS. As a compromise, HAF serves two NSG users and SG users, and the rest of NSG users are filled by the MBS. 
%Femtos that are configured in open access mode does not distinguish between SG users and non-SG users, and hence, they may fail to ensure QoS for SG users, especially during peak traffic loads.
Consequently, \textit{hybrid access} has been conceptualized as a trade-off between the two mechanisms. Hybrid Access Femtocells (HAFs) prioritize the QoS requirements of SG users over NSG users, by offering preferential access to radio resources to the former. This mechanism also simultaneously improves the overall capacity of the LTE HetNet by serving the NSG users in proximity of a HAF. Consequently, several \textit{rewarding} strategies have been put forward to make hybrid access mode attractive to the operators. However, the challenges of optimal placement of HAFs and efficient splitting of radio resources between SG and NSG indoor users need better solutions~\cite{New222,New333,New444,New555,New666}. \\
%To the best of our knowledge, none of the existing works discussed the fair allocation of radio resources among SG and NSG users to the best of our ability. This work proposes a dynamic bandwidth allocation method that divides the available bandwidth between the SG and NSG users and efficient power control and appropriate resource allocation method allocating the radio resources between SG and NSG users. 
 \item Improving Data Rates in LTE Small Cells and D2D Communication \\ \\
 Deployment of indoor Femtos significantly enhances network performance and capacity, but it also triggers a new set of problems. The resulting HetNet has to overcome the challenges of co-tier and cross tier-interference, frequent handovers, and signal leakage around the building's corners~\cite{sathya2016improving,comsnets}. The cross-tier interference in the HIZone close to the building adversely impacts the performance of outdoor UEs served by the one of the MBSs in the HetNet. Further, arbitrary or sub-optimal placement of Femtos often leads to high co-channel cross-tier interference among Femtos~\cite{rangisetti2020qos,sathyamodified}. It is also responsible for the existence of coverage holes in the indoor spaces. Moreover, lack of adequate power control mechanisms in Femtos leads to a surge in power consumption and also exacerbates inter-cell interference in large HetNets deployments. These problems are usually addressed through well-designed LTE small cell architecture, optimal Femto deployment, and smart power control to maintain a SINR threshold in Indoor environment. The D2D paradigm can also be leveraged by utilizing the idle IUEs connected to a Femto as relays to forward downlink data plane traffic to other UEs connected to MBS.\\
\item Asymmetric Vs. Symmetric ED threshold on LAA and Wi-Fi \\ \\
With the exponential rise in the number of mobile end-users, the mobile data demand on cellular and Wi-Fi infrastructure has also grown exponentially. Thus, it has become imperative that the licensed (cellular) and unlicensed (Wi-Fi) spectrum are utilized efficiently. Given the scarcity of spectrum for radio communication, cellular industry has actively pursued technological advancements that enable cellular networks to operate in the unlicensed spectrum as well. This, however, would require a fair coexistence of cellular technologies and Wi-Fi in the unlicensed spectrum. The IEEE 802.11 standard has prescribed an energy detection (ED) threshold of -62 dBm for Wi-Fi, whereas the LTE-LAA standard specifies that LTE-LAA nodes should detect Wi-Fi at -72 dBm. The impact of this asymmetry in the ED threshold has serious ramifications on fairness of LTE-WiFi coexistence~\cite{iqbal2017impact}. It has been demonstrated that lowering the Wi-Fi ED threshold from the prescribed -62 dBm improves performance of both Wi-Fi and LTE-LAA subsystems. Therefore, instead of treating LTE-LAA/LTE-U as noise, if Wi-Fi considers the coexisting LTE nodes as if they were overlapping Wi-Fi nodes, the performance of coexistence network improves significantly~\cite{iqbal2017impact}.\\

\item Facilitating LAA/Wi-Fi Coexistence Using Machine Learning Approach \\ \\
Several aspects of LTE-WiFi coexistence networks have been studied and analyzed in the current body of academic and industry research literature. However, analysis of real-world coexistence deployment data to study the performance and behavior of these deployments is acutely lacking. The problem of ``fair coexistence" of LTE-LAA and Wi-Fi, cell-selection in an LAA HetNet as compared to LTE HetNets, and performance prediction of LTE and LAA components in coexistence deployments are open probelms. Fortunately, with the fast roll-out of LAA deployments globally, LAA data can be collected and analyzed to analyze coexistence performance in the real-world. Some recent studies have adopted this approach and tried to address these problems. In~\cite{sathya2020measurement}, several issues of LAA-WiFi coexistence in real-world deployments of multiple cellular operators in downtown Chicago, have been explored. Aspects analyzed included the type of mobile traffic (e.g., data, video, and live streaming), extent of small cell coverage (Femto/LAA) coverage, and presence of secondary unlicensed carriers, among others. The study pointed out that a static channel allocation strategy of an LAA BS, leads to a particular channel being occupied for longer duration, adversely impacting the Wi-Fi APs as they face resource crunch due to the dynamic channel allocation strategy. An ML inspired channel assignment by analyzing the collected LAA data was proposed for the LAA BS and was shown to have a minimal adverse impact on Wi-Fi users. Likewise, deployment data of three LAA operators was analyzed to ascertain the impact of cell-selection mechanisms on LAA performance \cite{icdcn}. Further, a data-driven network hybrid optimization model for LAA networks was proposed in \cite{mobiquitous}.\\ \\
\item Association Issues in LTE-U/Wi-Fi Coexistence \\ \\
There are several challenges in ensuring association fairness in LTE-WiFi unlicensed coexistence in the same channel in the 5 GHz band. Beacon transmission is the initial step in the association process in  Wi-Fi, so the association fairness can be defined in terms of how fair LTE-U in letting Wi-Fi  transmit beacons on a shared channel that LTE-U has occupied through multiple duty cycles~\cite{sathya2018association,sathya2018analysis}.
  As per the LTE-U  specification, when the LTE-U  BS ascertains that the  channel is vacant, it can transmit for a duration up to  20ms and then turn OFF for only 1 ms, thereby resulting in a duty cycle of 95\%. However, a sizeable LTE-U duty cycle is likely to have an immense adverse impact on  the association fairness, especially during Wi-Fi beacon transmission and reception.  Studies have shown using NI USRP based experiments that a huge proportion of Wi-Fi beacons are neither transmitted in a timely manner nor are they received at the LTE-U BS~\cite{sathya2018association,sathya2018analysis}. An innovative Carrier Sense Adaptive Transmission (CSAT) approach was proposed in~\cite{sathya2020wi,manas2018socio,kala2020cirno} that aims to solve the challenges in Wi-Fi client association in a dense coexistence deployments while ensuring a fair spectrum access and sharing.\\

 \item Optimal Scaling of LTE-U Duty cycle in LTE-U/Wi-Fi Coexistence \\ \\
  Machine learning (ML) techniques offer efficient solutions to complex engineering problems and tasks such as video and image recognition, genome analysis, recommender systems, logistics, and automation of industrial operations. Given the potential of ML tools in solving non-linear problems, it is desirable to apply to the challenges encountered in spectrum sharing between unlicensed LTE (LAA/LTE-U) and Wi-Fi~\cite{sathya2019auto,adam2019detection,sathya2020machine,dziedzic2020machine,icdcn,mobiquitous,comsnets}. For example, LTE-U Forum has standardized the duty-cycle approach for fair coexistence  of LTE-U and Wi-Fi. However, the problem of identifying the precise number of APs by the LTE-U BS is not adequately addressed. If LTE-U BS can learn the exact numbers of Wi-Fi APs beforehand, it is possible to scale the LTE-U duty cycle optimally. However, this is a non-trivial problems as in order to detect the number of Wi-Fi APs operating on the same channel in real-time,  Wi-Fi packets need to be decoded which will require a Wi-Fi receiver at the LTE-U BS. Solutions to this problem has been put forward using innovative ML-based techniques which analyze power/energy levels observed in the channel during the LTE-U OFF duration. ML powered techniques offer higher accuracy when compared to the existing solutions based on auto-correlation (AC) and energy detection (ED)~\cite{sathya2018energy,kumar2019enhancing}.

\end{enumerate}

\section{Future NR-U Small Cell in 6 GHz}
 The current research (on Spectrum Sharing on 5 GHz) has opened many exciting possibilities to solve the research challenges for LTE/Wi-Fi coexistence in 6 GHz used as an unlicensed band~\cite{garg2019sla,kala2019odin}. In the following, we outline some of my future directions on the spectrum, sharing small cells.

\begin{figure}
\begin{center}
\includegraphics[width=0.5\textwidth]{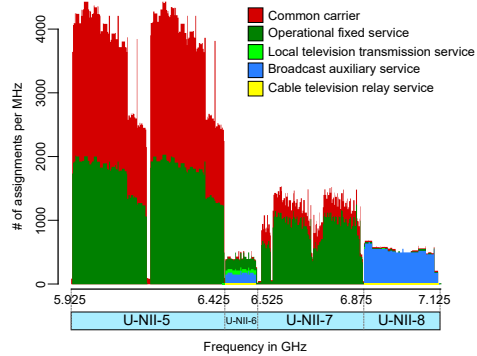}
%  \vspace{-0.5cm}
  \caption{Bandwidth Allocation on 6 GHz Spectrum \cite{FCC1}}
  \label{6ghz}
  \end{center}
\end{figure}

 \subsection{Fair Coexistence of NR-U and Wi-Fi in 6 GHz Spectrum~\cite{patriciello2020nr}:}
 
 Since the licensed spectrum is a limited and expensive resource, its optimal utilization may require spectrum sharing between multiple network operators/providers of different types. Increasingly licensed-unlicensed sharing is being contemplated to enhance network spectral efficiency beyond the more traditional unlicensed-unlicensed sharing. As the most common unlicensed incumbent, Wi-Fi is now broadly deployed in the unlicensed $5$ GHz band in North America, where approximately $500$ MHz of bandwidth is available. However, these $5$ GHz unlicensed bands are also seeing the increasing deployment of cellular services such as LTE-LAA. Recently, the Federal Communications Commission (FCC) sought to open up 1.2 GHz of additional spectrum for unlicensed operation in the 6 GHz~\cite{patriciello2020nr,coex} band through a Notice of Proposed Rule Making (NPRM) \cite{FCC1,WA,sathya2020standardization} as shown in Fig.~\ref{6ghz}.  Thus, this spectrum allocation for the unlicensed operation will only accelerate the need for other coexistence solutions among heterogeneous systems. Hence it is clear that regulatory authorities worldwide are paying close attention to the 6 GHz band as the next spectrum band that will continue to enhance unlicensed services across the world. However, it is also clear that this band, like the 5 GHz band, will see both Wi-Fi and cellular systems being deployed, and hence the coexistence issues played out in the 5 GHz band will repeat in this new frequency as well. In recognition of this, the two principal stakeholder standardization entities, IEEE and 3GPP, held a coexistence workshop in July 2019 \cite{coex} to discuss methods to address this before the next generation standards being specified. This section discusses the recent activities on FCC's 6 GHz NPRM and IEEE \& 3GPP efforts towards coexistence in the 6 GHz band.

\subsection{6 GHz Coexistence: Deployment Scenarios, and Channel Access}

\begin{figure}
\begin{center}
\includegraphics[width=0.5\textwidth]{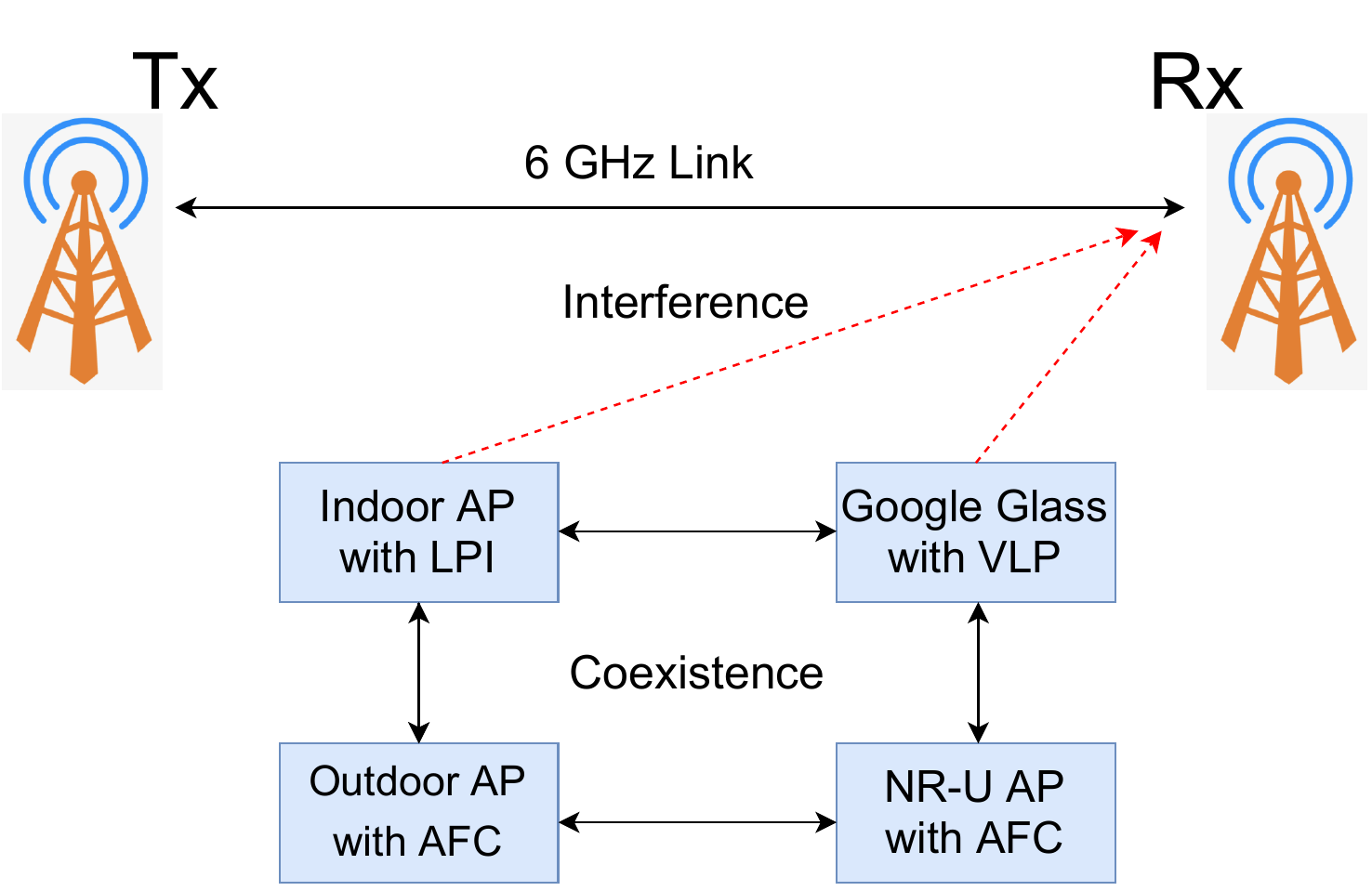}
%  \vspace{-0.5cm}
  \caption{Interference Scenario on 6GHz Spectrum\cite{FCC1}}
  \label{fig3}
  \end{center}
\end{figure}

Although several industry entities were not in favor of a re-evaluation, IEEE recommended that coexistence evaluations for NR-U should include 802.11ac (in 5 GHz), 802.11ax (in 6 GHz), and 802.11ad (in 60 GHz). Fig~\ref{fig3} shows the possible interference scenario in the 6GHz spectrum band, where the traditional licensed 6GHz link communication happens between transmitter and reception. But the different future use case applications (in 6GHz link) such as indoor AP with low power transmission (LPI), Google glass with very low power (VLP), outdoor AP with AFC database, and NR-U AP with database can potentially create the interference. For the sub-6 GHz bands, coexistence evaluations will be technology-neutral (e.g., channel access mechanism) and performed in random carrier frequencies in the 5 GHz band. These evaluations also necessitate devising suitable 11ac/ax coexistence topologies with a significant number of links below -72 dBm. 
\subsection{Future NR-U: Deployment Scenarios}
\begin{figure}
\begin{center}
\includegraphics[width=0.5\textwidth]{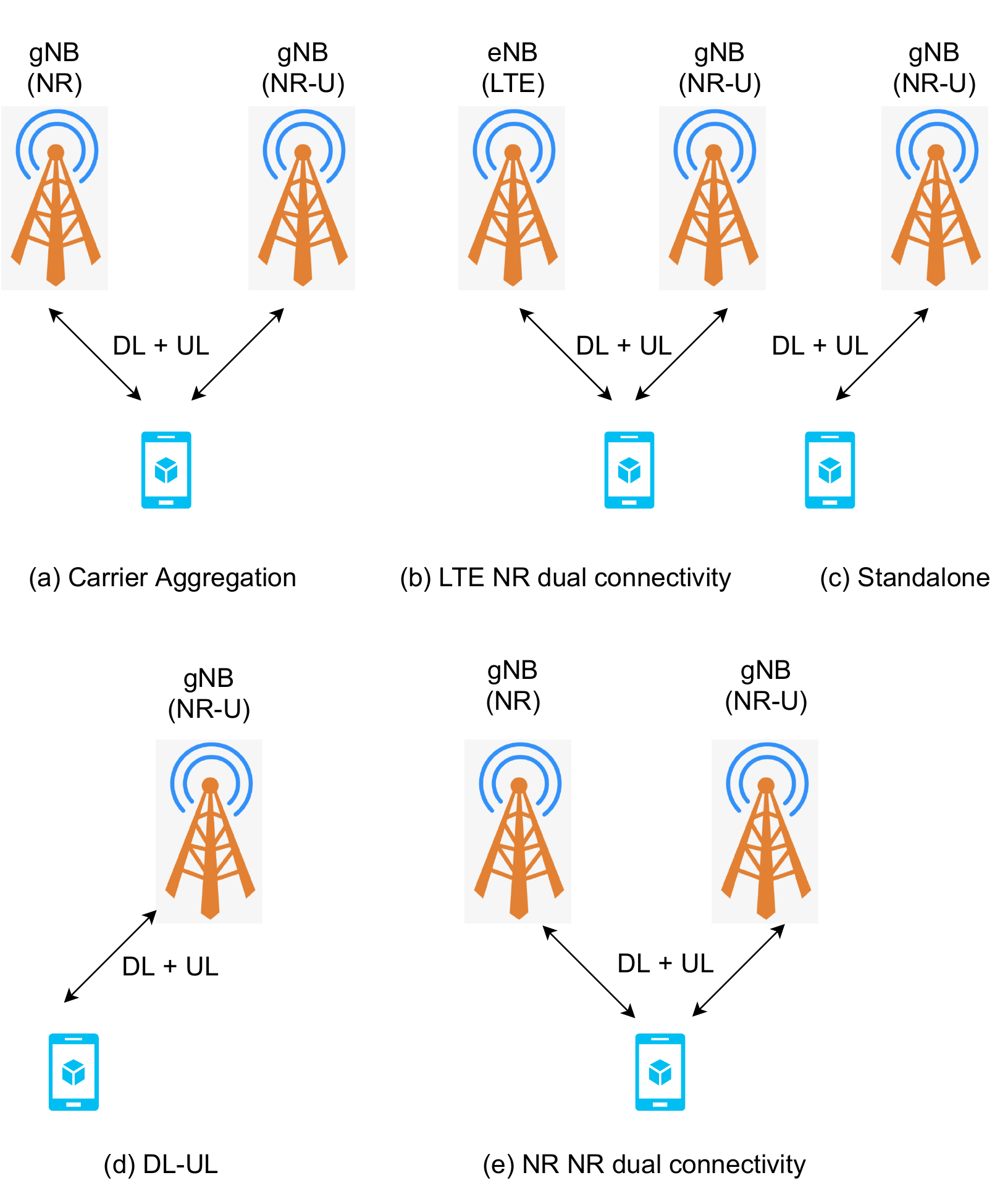}
%  \vspace{-0.5cm}
  \caption{NR-U Deployment Scenarios}
  \label{fig4}
  \end{center}
\end{figure}

The NR-U work item recently approved by 3GPP supports the existing unlicensed 5 GHz band and the new unlicensed "\textit{greenfield}" 6 GHz band. Industry players such as Qualcomm expect that other unlicensed and shared spectrum bands, including mmWave, will be added to this list in future releases. Researchers will study the following deployment scenarios to investigate the functionalities needed beyond the operation specifications in an unlicensed spectrum. The 3GPP forum propose different varieties of NR-U deployments as shown in Fig.~\ref{fig4}.
\begin{itemize}
\item \textit{Carrier aggregation} between licensed band NR (PCell) and NR-U (SCell): (a) NR-U SCell with both DL and UL. (b) NR-U SCell with DL-only.
\item \textit{Dual connectivity} between licensed band LTE (PCell) and NR-U (PSCell)
\item \textit{Stand-alone} NR-U
\item An NR cell with DL in the unlicensed band and UL in licensed band
\item \textit{Dual connectivity} between licensed band NR (PCell) and NR-U (PSCell)
\end{itemize}
The Legacy cellular operators oppose the NR-U stand-alone scenario and want 3GPP to drop it. They fear stiff competition from new players who can use NR-U stand-alone for limited cellular operation. NR-U is likely to be a more potent competitor to 802.11 than LAA as it will have a more flexible and efficient PHY/MAC marked by a shorter symbol duration, shorter HARQ Round Trip Time (RTT), etc. Further, NR-U can be deployed in every configuration where 802.11 is currently operational if both stand-alone and dual connection is approved. Also, unlike 802.11, NR-U will be capable of deploying the same PHY/MAC with flexible configurations across all current and future unlicensed bands.
\subsection{ML Based approaches to Solve Issues in Current and Future Spectrum Sharing:}

We have listed some of the interesting problems on LAA/Wi-Fi coexistence solved through an ML-based approach (as shown in Fig.~\ref{ml1}):

\begin{itemize}
    \item \textbf{Narrowband vs. Wideband LBT in 6 GHz:} The LBT mechanism is used by a device to avoid collisions by ensuring that no other transmissions are concurrently active in the channel. LTE-LAA follows CAT 4 LBT for most of its transmissions, while CAT 2 LBT is used for about 5\% of DL transmissions. NR-U is likely to adopt a mechanism similar to the LAA LBT. NR-U Release 16, like its predecessor, the NR Release 15, supports component carriers up to the maximum limit of 100 MHz bandwidth. Besides, it supports the aggregation of several inter and intraband component carriers. Multi-carrier LBT channel access as defined in 5 GHz is assumed \emph{i.e.,} the Type A LBT in 3GPP TS37.213, where each channel performs its independent LBT procedure. Consequently, there is bound to be high complexity when the operation bandwidth is wide. The alternative Type B LBT in 3GPP TS37.213 can reduce this complexity by performing a single LBT on multiple channels. The wideband LBT could simplify the implementation of wideband operation when it identifies that the channel is free of narrowband interference \emph{i.e.,} limiting the narrowband signal (20 MHz) certain sub-bands or by long/short term measurements and LBT bandwidth adaption.
Hence, wideband LBT is beneficial for systems operating with wide bandwidth as it simplifies LBT implementation.
    \item \textbf{Intelligent Selection of Unlicensed Channel by LAA BS:} While analyzing the collected data in real-time on LAA/Wi-Fi coexistence~\cite{sathya2020measurement}, we found that in an incredibly dense deployment scenario, multiple LAA operators contest for the unlicensed channel. However, numerous unlicensed channels are available, but choosing a particular channel and estimating the duration to occupy the track, with the vision of not affecting the Wi-Fi users and multiple LAA operators, is challenging. To solve this problem, we propose to use a Q-learning based ML solution so that a channel and its occupancy time decided intelligently based on the parameters such as interference from other operators, load on the channels, channel activities of Wi-Fi users, etc.
    \item \textbf{Intelligent Channel Selection by Wi-Fi AP in LTE/Wi-Fi Coexistence:} In a multi-AP setting, an AP selects the channel for the operation of the expected capacity of the existing links. The traditional way is to take the SINR-based capacity estimate into account. However, this capacity model may sometimes fail to represent the complex interactions between PHY and MAC layers due to the presence of LAA, as it makes the scenario heterogeneous. As a result, decisions regarding channel selection may be delayed or inaccurate. To solve this problem, we propose to use supervised learning as a tool to model the complex interactions between PHY and MAC layers based on factors such as power and PHY rate of a neighboring Wi-Fi link.
    \begin{figure}
\begin{center}
\includegraphics[width=0.5\textwidth]{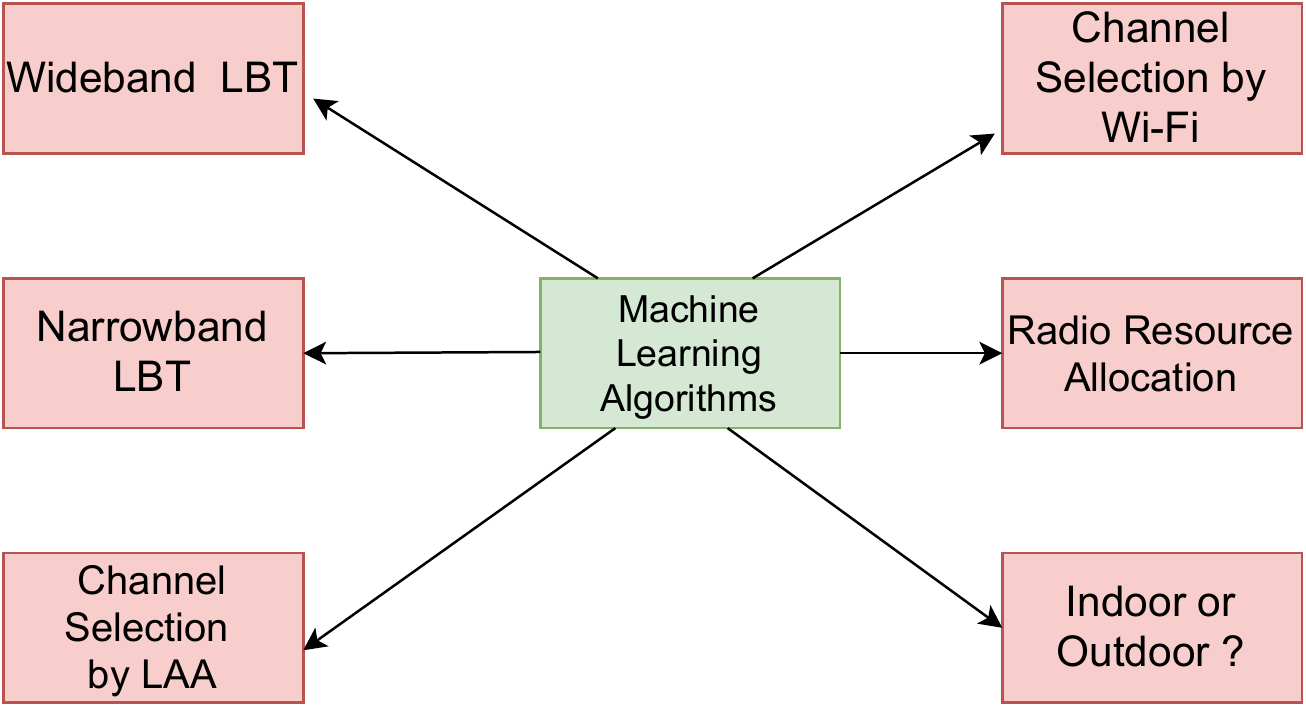}
%  \vspace{-0.5cm}
  \caption{Machine learning based approaches in spectrum sharing }
  \label{ml1}
  \end{center}
\end{figure}

    \item \textbf{Efficient Radio Resource Allocation Using Reinforcement Learning:} In the fifth-generation (5G) of mobile broadband band systems, Radio Resources Management (RRM) has reached unprecedented levels of complexity. To cope with the ever more sophisticated RRM functionalities and to make prompt decisions required in 5G, efficient radio resource scheduling will play a critical role in the RAN system. Depending on tasks' purposes, the scheduling process is divided into three steps: prioritization, resource number determination, and resource allocation. However, to support a diverse range of applications such as ultra-reliable low latency applications, IoT applications, V2X applications, AR/VR applications, massive multimedia applications, etc., the scheduling process's steps become more complex. As a solution, we propose to use deep reinforcement learning tools to get feedback about the traffic in real-time so that efficient scheduling decisions and optimal use of radio resources are made. 
    \item \textbf{Indoor and Outdoor classification for spectrum sharing:}  The specifications for 6 GHz  \cite{FCC1} provide two different power regimes for unlicensed UEs. Devices within the built environment or IUEs are expected to stick to lower transmit powers, but do no have to seek permission to use a channel by first accessing an Automatic Frequency Control (AFC) database. On the other hand,  outdoor devicesor OUEs, are allowed to transmit at higher power levels, but are required to seek permission through AFC access prior to transmitting on a channel, so as to ensure that the OUE is not excluded from the particular channel. However, determination of a UEs location, \emph{i.e.,} whether it is indoors or outdoors is a non-trivial problem. Therefore, specifications prescribed a few more restrictions. UEs which are able to connect to an indoor AP but could be situated outdoors are subjected to the constraint of a 6 dB lower transmit power threshold. Further, indoor APs can not be battery powered, must be equipped with detachable antennas, or should have a weatherized exterior. However, these constraints result in a sub-optimal spectrum efficiency. For example, an IUE always has to transmit at a lower power. Further, UEs are not permitted to communicate directly with each other without first connecting through a Wi-Fi AP. More importantly, not permitting APs to be battery powered erodes network resilience and robustness. Thus, it is of extreme importance that the environment of a wireless device is reliably detected so as to ensure optimal spectrum utilization, enhanced resilience, and improved system capacity. It is noteworthy that the ambient RF environment, differs markedly in an indoor setting as compared to outdoors, in terms of the signal strength. This is true for all commercially used communications bands, such as 2.4 GHz Wi-Fi, 5 GHz Wi-Fi, 6 GHz Wi-Fi (in the future), several low cellular ($<$ 1 GHz), mid cellular (1 GHz - 6 GHz) and high cellular ($>$ 24 GHz) bands. As highlighted earlier, measuring the signal strengths in these different environments and analyzing the data using ML algorithms, intelligent models and methods for the prediction of device-environment can be designed. Doing so will enhance spectrum usage and offer security and resilience, not only in the 6 GHz, but also in future bands such as the 12 GHz satellite band.
\end{itemize}

\section{Conclusions}
A few important conclusions and inferences can be drawn from the discussion presented in this work. Clearly, the best way to improve spectrum efficiency, and in turn network performance, is by intelligently reusing the scarce radio bands to the maximal extent possible. Deploying small cells is one way to achieve the \textit{spatial reuse} of spectrum. Thus, Pico cells and Femto cells are deployed in indoor and outdoor settings and are designed to optimize transmission power and perform interference management through techniques such as Full reuse, Soft reuse, Hard reuse, and Strict reuse. Small cell deployments can also leverage the state-of-the-art antenna technology at their disposal such as directional and omni-directional antennas. With the the help of these technologies, small cell deployments in LTE/5G are vital to delivering high capacity and fault-tolerant networks to the end-user. Further, this work outlined several open challenges and problems in the current (5GHz) and the future (6GHz, NR-U) unlicensed spectrum sharing. Rapid proliferation and adoption of latest Wi-Fi standard, \emph{i.e.,} 802.11ax, and the latest cellular technology, \emph{i.e.,} 5G NR-U, is expected. Further, unlike LAA, 5G NR-U will be capable of transmitting data in both uplink and downlink in the unlicensed band. Likewise, unlike 802.11ac in 5 GHz, 802.11ax in 6 GHz will employ Orthogonal Frequency Division Multiple Access (OFDMA). Together, 802.11ax and 5G NR-U coexistence will spawn new challenges and problems in unlicensed coexistence. Thus, all future research in unlicensed coexistence is likely to be focused on fair and efficient LTE-WiFi coexistence deployments in the 6 GHz unlicensed band.

%\vspace{-0.5cm}
%\bibliographystyle{elsarticle-num}

\end{document}